\documentclass[twocolumn,5p,times]{elsarticle}

\usepackage{amssymb}

\usepackage{lineno}

\usepackage[usenames]{color}

\usepackage{nicefrac}
\usepackage{multirow} 
\usepackage{hyperref}
\usepackage{units}
\usepackage{fixltx2e}  % keeps the order between figure and figure* environments
\usepackage{dblfloatfix}
\usepackage{amsmath}
\usepackage{bm} % bold greek letters and symbols

\journal{Astroparticle Physics}

\begin{document}

\begin{frontmatter}

\title{Reconstruction of air shower muon densities using segmented counters with time resolution}

\author[ITEDA]{D. Ravignani \corref{cor1}}
\cortext[cor1]{Corresponding author: diego.ravignani@iteda.cnea.gov.ar}
\author[IAFE]{A. D. Supanitsky}
\author[ITEDA]{D. Melo}
\address[ITEDA]{ITeDA (CNEA, CONICET, UNSAM), Buenos Aires, Argentina.}
\address[IAFE]{Instituto de Astronom\'{\i}a y F\'{\i}sica del Espacio (IAFE, CONICET-UBA), Buenos Aires, Argentina.}

\begin{abstract}
Despite the significant experimental effort made in the last decades, the origin of the ultra-high energy cosmic rays is still largely unknown. 
Key astrophysical information to identify where these energetic particles come from is provided by their chemical composition.
It is well known that a very sensitive tracer of the primary particle type is the muon content of the showers generated by the interaction of the cosmic rays with air molecules. 
We introduce a likelihood function to reconstruct particle densities using segmented detectors with time resolution.
As an example of this general method, we fit the muon distribution at ground level using an array of counters like AMIGA, one of the Pierre Auger Observatory detectors.
For this particular case we compare the reconstruction performance against a previous method. 
With the new technique, more events can be reconstructed than before.
In addition the statistical uncertainty of the measured number of muons is reduced, allowing for a better discrimination of the cosmic ray primary mass. 
\end{abstract}


\end{frontmatter}

%\linenumbers

\section{Introduction}

% The composition problem
Although the origin of the ultra-high energy cosmic rays is still unknown, significant progress has been recently achieved from data collected by setups like the Pierre Auger Observatory~\cite{Aab:2015zoa} and the Telescope Array~\cite{AbuZayyad:2012kk}. 
The three main observables used to study the nature of cosmic rays are their energy spectrum, arrival directions, and chemical composition. 
Certainly, composition is a crucial ingredient to understand the origin of these very energetic particles~\cite{Kampert:2012mx}, to find the spectral region where the transition between the galactic and extragalactic cosmic rays takes place~\cite{MedinaTanco:2007gc}, and to elucidate the origin of the flux suppression at the highest energies~\cite{Kampert:2013dxa}.

% Motivation for counting muons
For energies larger than $10^{15} \,\mathrm{eV}$, cosmic rays are studied by observing the atmospheric showers produced when they interact with the air molecules. 
Therefore composition has to be inferred indirectly from parameters measured in air shower observations. 
The observables most sensitive to the primary mass are the depth of the shower maximum and the number of muons generated during the cascade process. 
While the maximum depth is observed with fluorescence telescopes, the muons are measured at ground level and underground with surface and buried detectors respectively.
Besides composition, hadronic interactions can also be studied with muons.
At the highest cosmic ray energies the hadronic interactions are unknown, so models that extrapolate accelerator data at lower centre-of-mass energy are used in shower simulations. 
As the number of muons predicted by simulations strongly depends on the assumed interaction model, the muon data can be used to discriminate among different scenarios~\cite{Supanitsky:2008ph,Aab:2014dua,Aab:2014pza,Apel:2014qqa,Apel:2009sv}.

%AMIGA
In Auger, using the water-Cherenkov detectors of its surface array, muons have been measured by disentangling them from other shower particles.
However this technique can only be applied when muons produce a large fraction of the total signal. Those special cases include inclined showers with zenith 
angle between $62^{\circ}$ and $80^{\circ}$~\cite{Aab:2014pza}, and also showers close to $60^{\circ}$. However, in this second case, only detectors 
more than $1000 \,\mathrm{m}$ away from the shower core are used~\cite{Aab:2014dua}. To include the more abundant vertical showers and to extend 
the reach to lower energies, dedicated muon counters are called for. Currently Auger is building a triangular array of muon counters spaced every 
$750\,\mathrm{m}$ as part of the AMIGA project~\cite{Wundheiler:2015}. Once finished the AMIGA array will cover $23.5\,\mathrm{km^2}$ in a small 
region of the surface detector. The detector is designed to measure showers between $3 \times 10^{17} \,\mathrm{eV}$ and $10^{19} \,\mathrm{eV}$, 
the upper limit determined by the number of events that can be collected given the detector size. Each grid location will have three $10\,\mathrm{m^2}$ 
counters made out of plastic scintillator, buried $2.5\,\mathrm{m}$ underground, and divided into 64 scintillator strips of equal size. The three counters installed at 
each array site are equivalent to a single $30\,\mathrm{m^2}$ detector divided into $192$ bars. Muons are counted in time windows of $25\,\mathrm{ns}$, 
the duration corresponding to the detector dead time given by the width of the muon pulse. 

% Punch-through
Close to the shower core the muons are accompanied by energetic electrons and gammas. 
However the soil shielding significantly reduces the contamination of the detector signals by these electromagnetic particles.
The soil density at the AMIGA site, $2.4$ g cm$^{-3}$, entails a shielding of $22$ radiation lengths at $2.5\,\mathrm{m}$ underground. 
Using these parameters, shower simulations including the propagation of particles underground show that the electromagnetic contamination is negligible in AMIGA but very close to the shower core~\cite{PierreAugur:2016fvp}.

% Setting the context for our likelihood
AMIGA measures the fall of the muon density with the distance to the shower axis, i.e. the so-called \emph{lateral distribution function} (LDF). 
The LDF evaluation at a reference distance is a long-established method to characterise the size of an air shower~\cite{Newton:2006wy}.
In the surface arrays of the cosmic ray observatories, the LDF is fitted to the detector data by either minimising a $\chi^2$ or by maximising a likelihood function~\cite{Apel:2010zz, Nagano:1984eb}.
The used likelihood, modelling the detector response to incoming particles, is specific to each detector type.
In this paper we present a likelihood suitable for a particular detector, namely a segmented particle counter with time resolution like that used in AMIGA.

% AMIGA previous likelihoods
We fit the LDF to the detector data by maximising a likelihood that links a muon density to the observed signals.
We previously used two likelihood models. 
In the first method we adopted an approximation valid for few muons in a detector~\cite{Supanitsky:2008dx}. 
Using this approach we showed in~\cite{Ravignani:2014jza} that detectors saturate if there are more than 174 muons in a time window.
As consequence events with a core falling less than $100\,\mathrm{m}$ from a detector cannot be reconstructed.
To enlarge the statistics we later proposed another likelihood model valid for higher signals, thus covering an interval where the detector response departs from linearity. 
In this second case, to obtain an analytic expression, the time resolution of the detector had to be neglected. 
This method just considered whether a scintillator bar has a signal during the whole duration of the event. 

% The new likelihoods
Although the second likelihood improved the original one, grouping muons into a single time window is a drawback since shower particles arrive at the ground spread in time. 
For both the electromagnetic and muonic shower components, the Kascade-Grande array has measured signal widths of \unit[70]{ns} beyond \unit[400]{m} from the core~\cite{Antoni:2001pi}. 
At larger core distances, common in larger observatories, the particles arrive even more widespread and, consequently, the air shower signals extend over many \unit[25]{ns} time windows.
To make the best use of the detector capabilities, we improved the likelihood by including the signal timing.
We started by considering the complete likelihood of a segmented detector with time resolution. 
To get rid of nuisance parameters present in the full likelihood, we applied two different approximations: the profile~\cite{Agashe:2014kda} and the integrated likelihoods~\cite{Berger:99}. 
The first technique, well established in the field of high-energy physics, was used in the discovery of the Higgs boson~\cite{Aad:2015zhl}. 

% Article structure
The following section describes the profile and the integrated likelihoods, and section~\ref{sec:examples} illustrates them with examples.
Section~\ref{sec:simulations} presents the simulations used to evaluate the likelihoods. 
We compare the performance of the new and old methods in section \ref{sec:performance}, and conclude in section~\ref{sec:conclusions}.   

\section{Likelihood of a segmented detector}
\label{sec:likelihood}

\subsection{Likelihood of a single time bin}
\label{ssec:likelihood1}

% Present the statistics (number of segments with a signal)
We built the profile and integrated likelihoods as extensions of the single-window likelihood developed in~\cite{Ravignani:2014jza}. 
For completeness some of the  material developed in that work is summarised below. 
We must recall that the main goal of the counters used in a cosmic ray observatory is to estimate a particle density ($\rho$).
The density multiplied by the detector area ($a$) and the zenith angle cosine of the shower direction is the average number of particles expected in the counter ($\mu$),
\begin{equation}
\mu  =  \rho \, a \, \cos\theta.
\label{eq:mu}
\end{equation} 
\noindent In turn, $\mu$ is the parameter of a Poisson distribution that describes the actual number of particles impinging on the detector.
Correspondingly, for a detector divided into $n$ parts, the number of muons in each segment fluctuates according to a Poissonian  with parameter $\mu/n$.

The arriving particles produce a signal in some of the detector segments.
Occasionally two or more muons pile up in the same segment. 
Depending on the number of particles, each segment can take two distinct states: \emph{on} if hit by one or more muons, and \emph{off} otherwise.
According to Poisson, the probability of a segment \emph{off} is $q =  e^{-\nicefrac{\mu}{n}}$, and the odds of an \emph{on} state is $p = 1-q$.
Since the segment states are independent from each other, the probability of $k$ segments \emph{on} out of a total of $n$ segments follows the binomial distribution,
\begin{equation}
P(k ; \mu)  = L(\mu ; k) = {n \choose k} \, p^k \,q^{n-k} = {n \choose k} \, e^{-\mu} \, \left( e^{\nicefrac{\mu}{n}} -1 \right)^k.
\label{eq:kprob}
\end{equation}
\noindent In addition to a probability, Eq.~(\ref{eq:kprob}) is the likelihood of $\mu$ expected muons when $k$ strips out of $n$ are \emph{on}. If $k<n$, the corresponding maximum likelihood estimator ($\hat{\mu}$) is,  
\begin{equation}
\hat{\mu} = -n \, \ln\left(1-\frac{k}{n}\right). 
\label{eq:muest}
\end{equation}
%
% Saturation
\noindent If $k=n$ the likelihood tends to unity when $\mu$ increases, and the maximum likelihood estimator of $\mu$ tends to infinity.
In this case, the likelihood sets a lower bound to the number of muons allowed in the LDF fit~\cite{Ravignani:2014jza}. 
Based on this behaviour we labelled these detectors as \emph{saturated}.    

% Model assumptions
The proposed likelihood only considers the detector size and segmentation.
This function excludes any signal contamination produced either in the detector electronics or in the photomultipliers.
This simplified model of the likelihood is realistic because the AMIGA detector filters out the detector noise. 
The electronic noise is filtered by tuning the discrimination level applied to the analogue signals produced by the photomultipliers.
In turn any casual photomultiplier after pulse is removed by requiring the digital signals to be compatible with at least two photoelectrons~\cite{Wundheiler:2011zz}.  

\subsection{Profile likelihood}
\label{subsec:proflike}

To extend the likelihood to many time bins, one has to consider the time spread of the muon signal $d\mu(t)/dt$. 
The number of expected muons ($\mu$) is the integral of this signal over the event duration, $\mu = \int \frac{d\mu(t)}{dt} \, dt$.
Correspondingly, within a time bin, the number of muons ($\mu_i$) is the integral restricted to the window limits. 
The sum of the $\mu_i$'s is $\mu$. 

% describe the muon counting algorithm
The AMIGA segmented detector counts particles in windows of $25\,\mathrm{ns}$.
For each of these time bins, the number of strips \emph{on} ($k_i$) is computed.
Considering that the $k_i$'s of different time windows are independent from each other, the likelihood of $\mu_i$ particles in the $i$-th bin is given by Eq.~(\ref{eq:kprob}).
The likelihood of all time bins $\big(L(\boldsymbol{\mu})\big)$ is the product of the single-window likelihoods, 
\begin{equation}
  L(\boldsymbol{\mu}) = \prod_{i=1} L_i(\mu_i),  
  \label{eq:likelihood}
\end{equation} 
\noindent where $i$ runs over the time bins and $\boldsymbol{\mu} = (\mu_1,\mu_2,\dotsc)$.

% Profile likelihood
In the LDF fit, the parameter of interest is the total number of muons $\mu$.
However the value of $\mu$ alone is not enough to calculate the likelihood because this function also depends on each of the $\mu_i$'s.
An obstacle arises at this point, the lack of knowledge of the signal time distribution $d\mu(t)/dt$ prevents us from deriving the $\mu_i$'s from $\mu$.
We overcame this issue by using a profile likelihood ($L_P(\mu)$). 
Following this approximated method we searched, for each $\mu$, the likelihood maximum under the restriction $\sum \mu_i = \mu$,
\begin{equation}
 L_P(\mu) = \max_{\sum \mu_i = \mu} L(\boldsymbol{\mu}). 
\end{equation}

%Profile likelihood ratio
In this treatment of the likelihood, the $\mu_i$'s are nuisance parameters which are fixed by applying the profiling technique.
We performed the likelihood maximisation with the Minuit library~\cite{James:1975dr} implemented in the ROOT data analysis framework~\cite{Antcheva:2009zz}.
For some desirable mathematical properties mentioned below, we used the \emph{profile likelihood ratio} defined as,
\begin{equation}
\lambda(\mu) = \frac{L_P(\mu) }{ L_{\mathrm{max}} }.
\label{eq:likelihooRatio}
\end{equation} 
\noindent where $L_{\mathrm{max}}$ is the global maximum of the likelihood calculated without any restriction on $\boldsymbol{\mu}$.
The likelihood reaches this maximum when $\boldsymbol{\hat\mu}=(\hat{\mu}_1,\hat{\mu}_2,\dotsc)$, all given by Eq.~(\ref{eq:muest}).
From Eq.~(\ref{eq:likelihooRatio}) one can see that $\lambda$ varies between 0 and 1, the maximum value attained at,
\begin{equation}
\hat\mu = \sum \hat{\mu}_i.
\label{eq:muest2}
\end{equation} 
\noindent A $\lambda$ close to unity means a likely value of $\mu$ given the observed data, i.e. a $\mu$ close to $\hat\mu$.
On the other hand, a low $\lambda$ implies an unlikely $\mu$.

\begin{figure} [tbh]
\centering
\setlength{\abovecaptionskip}{0pt}
\includegraphics[width=.45\textwidth]{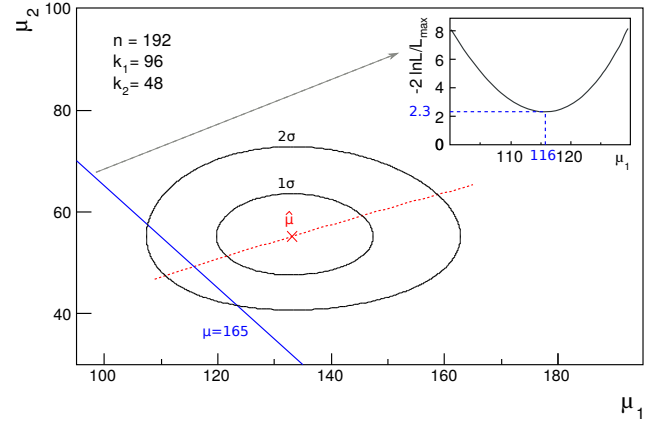}
\caption{Contour levels of the function $-2\,\ln L(\boldsymbol{\mu}) / L_{\mathrm{max}}$ for a signal spread over two time bins.
The parameters $\mu_1$ and $\mu_2$ are the numbers of muons in each bin.  
In this example the detector is divided into $n=192$ segments, the first time bin has $k_1=96$ bars \emph{on}, and the second one $k_2=48$.
The red cross indicates the global maximum $\boldsymbol{\hat\mu}$ of the likelihood $L(\boldsymbol{\mu})$ and the dotted red line the corresponding local maxima at constant $\mu=\mu_1+\mu_2$.
Two contour levels defining the $\sigma$ standard-deviation regions of $\boldsymbol{\mu}$~\cite{Agashe:2014kda} are displayed.
The continuous blue line corresponds to a cut at a sample $\mu=\mu_1+\mu_2=165$.
{\bf Inset}: Function $-2\,\ln L(\boldsymbol{\mu}) / L_{\mathrm{max}}$ along the cut $\mu=165$. The local minimum is reached at $\mu^{*}_1=116$. 
\label{fig:like2b}}
\end{figure}

% The statistics -2lnλ
Providing certain conditions are met, the distribution of $f(\mu)=-2\ln\lambda(\mu)$ approaches a $\chi^2$ distribution, independently from the nuisance parameter values~\cite{Wilks:1938}.
For a segmented particle counter, these requirements translate to having many muons. 
However the number of particles must not be so high as to saturate the detector.
An upper limit to the number of muons is approximately three times the number of detector strips. 
This bound corresponds to the probability of a segment \emph{on} to be 0.95. 
In most formal terms, this condition is equivalent to asking that the binomial distribution of the window with more muons can be approximated by a Gaussian.  
Considering the values taken by $\lambda$, $f(\mu)$ is always positive and drops to zero at $\hat{\mu}$.
If the quoted asymptotic conditions are met, $f(\mu)$ is approximately quadratic in a wide region around $\hat{\mu}$.
Correspondingly, in the LDF fit, the detector $f(\mu)$ is equivalent to a $\chi^2$ with a $\sigma$ given by the width of the likelihood. 
%
% Example: profile likelihood procedure 
The procedure to obtain the profile likelihood is illustrated in Fig.~\ref{fig:like2b} with a signal spread over two time windows.

\subsection{Integrated likelihood}

Besides the profile likelihood, another useful technique to get rid of nuisance parameters is the integrated likelihood. 
While in the profile technique the nuisance parameters that maximise the likelihood are searched for, in this second method the likelihood is integrated over these parameters. 
To introduce the integrated likelihood, let us first rewrite the nuisance parameters as $p_i=\mu_i/\mu$. 
Consequently the condition $\sum \mu_i = \mu$ is now given by $\sum p_i = 1$. 
Considering this restriction and the single bin likelihood of Eq.~(\ref{eq:kprob}), the integrated likelihood can be written as, 
\begin{eqnarray}
\label{eq:IntL}
\begin{aligned}
L_I (\mu) \propto & \int_0^1 dp_1 \cdots \int_0^1 dp_N \prod_{i=1}^N\exp(-\mu\ p_i)   \\
& \times \left( \exp(\mu\ p_i/n)-1 \right)^{k_i} \delta\left(\sum_{i=1}^N p_i - 1 \right),
\end{aligned}  
\end{eqnarray} 
\noindent where $N$ is the number of time bins and $\delta(x)$ is the Dirac delta function. 

In most cases, the integral in Eq.~(\ref{eq:IntL}) has to be calculated numerically, however for the case of two time intervals an analytic expression can be obtained (see Sec.~\ref{TwoBins}). 
The integrated likelihood requires the calculation of multidimensional integrals which we computed
using the VEGAS algorithm \cite{Lepage:78} implemented in ROOT.
The computation of many time bins takes a long time; so we reduced the number of involved integrals by calculating all the intervals having the same $k_i$ with a single integral.
Applying this optimisation (see \ref{App} for details), we arrived to the following approximated expression of the integrated likelihood,
\begin{equation}
\begin{aligned}
\label{IntLm}
L_I (\mu) \propto & \int_0^{1/m_1} dp_1 \cdots \int_0^{1/m_{\tilde{N}}} dp_{\tilde{N}} \prod_{i=1}^{\tilde{N}} \exp(-\mu\ p_i\ m_i) \\
& \times \left( \exp(\mu\ p_i/n)-1 \right)^{k_i\ m_i}\ p_i^{m_i-1} 
\ \delta\left(\sum_{i=1}^{\tilde{N}} p_i\ m_i - 1 \right),   
\end{aligned}  
\end{equation} 
\noindent where $m_i$ is the multiplicity of the $k_i$ value and $\tilde{N}$ is the number of $k_i$ values that are different among them.  

\section{Likelihood examples}
\label{sec:examples}

\subsection{The few muons limit}

% Why doing this is important...
So far we presented the complete likelihood of a segmented detector and two different approximations applied to get rid of nuisance parameters.
It is a desirable mathematical property that, in some limiting case, the approximations and the full method converge to the same function.
This condition is met by the three introduced likelihoods if the number of muons is small compared to the number of detector segments; in this case all of them tend to a Poisson distribution.
Below we calculate this limit for each method. 

% The case of the complete likelihood
The demonstration for the full likelihood starts with the single-window likelihood of Eq.~(\ref{eq:kprob}).  
If $\mu_i \ll n$, the binomial distribution of $k_i$ can be approximated by a Poisson distribution with parameter $\mu_i$. Then the distribution of the variable $k=\sum k_i$ follows a Poissonian with parameter $\mu=\sum \mu_i$. The corresponding likelihood is,
\begin{equation}
  L(\mu) = e^{-\mu} \, \frac{\mu^k}{k!}.  
  \label{eq:lpoiss}
\end{equation} 

The function of Eq.~(\ref{eq:lpoiss}) does not depend on the individual nuisance parameters $\mu_i$ but on their sum, i.e. the likelihood is profiled.  
For the integrated likelihood, the independence of the distribution on the nuisance parameters $p_i$ allows the extraction of the integrand in Eq.~(\ref{eq:IntL}) to arrive to, 
\begin{eqnarray}
\label{IntLPoiss}
L_I (\mu) \!\!\! &\cong& \!\!\! \exp(-\mu)\ \mu^{k} \int_0^1 dp_1 \cdots \int_0^1 dp_N \prod_{i=1}^N 
\left( \frac{p_i}{n} \right)^{k_i}  \nonumber \\ 
&& \!\!\! \times\ \delta\left(\sum_{i=1}^N p_i - 1 \right), \nonumber \\ 
\!\!\! &\propto& \!\!\! \exp(-\mu)\ \mu^{k}, 
\end{eqnarray} 
\noindent which corresponds to a Poisson likelihood. 
One has to consider that the Poisson approximation is only valid in the limited range of small $\mu$. 
In the fit, the approximation must hold for likely values of $\mu$, i.e. the region around the likelihood maximum $\hat\mu$.
In terms of the data, this condition is equivalent to asking that, via Eq.~(\ref{eq:muest}), $k_i \sim \mu_i \ll n$.
Therefore, if the number of segments \emph{on} is small compared to the detector segmentation, the exact, the profile, and the integrated likelihoods are well approximated by the same Poisson function.

\subsection{Example for two time bins \label{TwoBins}}

%The case of two windows
The evaluation of $f(\mu)=-2\ln\lambda(\mu)$ requires a numerical minimisation to calculate the profile likelihood. 
However, in the special case of only two time windows, $f(\mu)$ has the analytic expression,
\begin{equation}
\begin{aligned}
 f(\mu) &= 2 \, \mu - 2 \sum_{i=1,2} \bar\delta[k_i] \, \bigg( k_i \ln(k_i/n)  \\
 &\quad  + (n-k_i) \ln(1-k_i/n) - \, k_i \ln \left( e^{\mu^{*}_i/n}-1 \right) \bigg),  
\end{aligned}
\label{eq:like2b}
\end{equation}

\noindent where $\mu^{*}_1$ and $\mu^{*}_2$ are the number of muons in each time window. 
These values correspond to the local maximum of the likelihood at constant $\mu_1+\mu_2=\mu$.
The function $\bar\delta[k_i]$, used to include the case of $k_i=0$, is 
zero at $k_i=0$ and one otherwise. The value of $\mu^{*}_1$ is,
\begin{equation}
  \mu^{*}_1(\mu) =
\begin{cases} 
 -n \, \ln \left( -\frac{k_1-k_2}{2\,k_2} + \sqrt{ \left( \frac{k_1-k_2}{2\,k_2} \right)^2 + \frac{k_1}{k_2} \, e^{-\mu/n} } \, \right) & \text{if } k_1 > 0 \\
 0 & \text{if } k_1 = 0.
 \label{eq:alpha} 
\end{cases} 
\end{equation}

\noindent Correspondingly $\mu^{*}_2$ is $\mu-\mu^{*}_1$.
The function $f(\mu)$ depends on $\mu$ explicitly as per Eq.~(\ref{eq:like2b}) and also indirectly through the $\mu^{*}_i$'s. 
If $k_1=k_2$, it can be seen from Eq.~(\ref{eq:alpha}) that $\mu^{*}_1=\mu^{*}_2 = \mu/2$.
We exploited this degeneracy, also present in the general case of more than two time bins, to reduce the number of nuisance parameters. 
By using fewer free parameters, we optimised the numerical minimisation run to evaluate the profile likelihood.

Also for the integrated likelihood technique it is possible to find an analytic expression of the likelihood as a function of $\mu$ which is given by,
\begin{equation}
\begin{aligned}
L_I (\mu) =& \exp(-\mu)\ \sum_{i=1}^{k_1}  
\sum_{j=1}^{k_2}  {{k_1}\choose{i}}\ {{k_2}\choose{j}}\ (-1)^{k_1+k_2-i-j}  \\
& \times \xi(\mu,i,j,n),
\end{aligned}
\end{equation}
where,
\begin{equation}
%\begin{aligned}
\xi(\mu,i,j,n) = \left\{ 
\begin{array}{ll}
\exp(\mu\ j/n) & i=j\\
& \\
n\ \mathop{\displaystyle \frac{\exp(\mu\ i/n)-\exp(\mu\ j/n)}{\mu\ (i-j)} } & i\neq j 
\end{array}    \right.. 
%
%\end{aligned}
\end{equation}

\begin{figure} [tbh]
\centering
\setlength{\abovecaptionskip}{0pt}
\includegraphics[width=.45\textwidth]{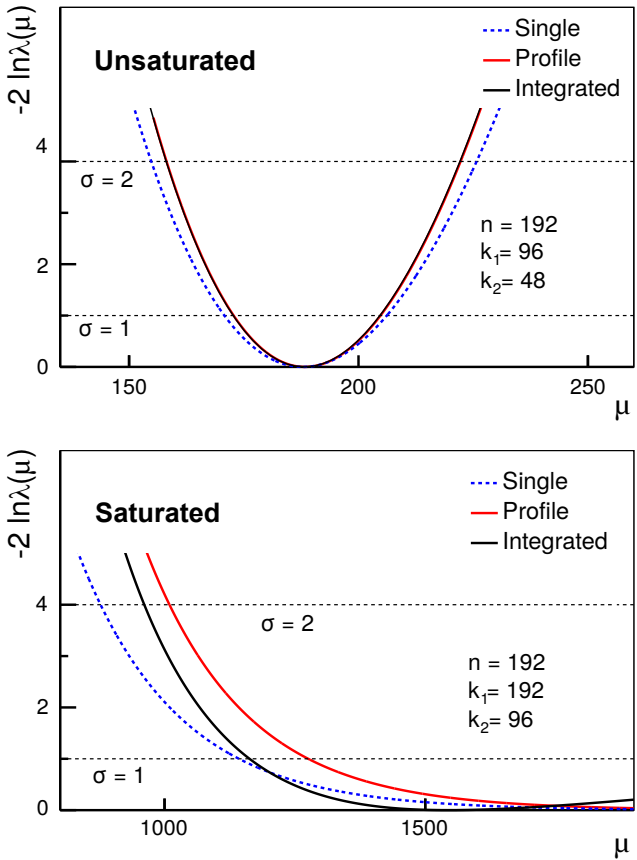} 
\caption{Single-window, profile and integrated likelihoods for a detector divided into $192$ segments. 
The parameter $\mu$ is the number of muons in the detector. 
Top: Counter with $k_1=96$ and $k_2=48$ segments \emph{on} in the first and second time bins respectively (same example of Fig.~\ref{fig:like2b}). For the single-window likelihood we assumed $k=120$ segments \emph{on}.
Bottom: Saturated detector with $192$ and $96$ bars \emph{on} in the first and second time bins, respectively. \label{fig:like1}}
\end{figure}

We show next a comparison of the likelihoods corresponding to the two-window example of section~\ref{subsec:proflike}.
The dotted red line in Fig.~\ref{fig:like2b} shows the local maxima of the likelihood $L(\boldsymbol{\mu})$ at different values of $\mu$.
The likelihood is evaluated along this curve to calculate $f(\mu)=-2\ln\lambda(\mu)$ via the profile likelihood.   
The $f(\mu)$ corresponding to the single-window, profile, and integrated likelihoods are shown in the top panel of Fig.~\ref{fig:like1}.
The maximum likelihood estimator of the number of muons is $\hat\mu=\hat\mu_1+\hat\mu_2=188.3$ for both the profile and the integrated likelihoods. 
The number of strips \emph{on} required in single-window likelihood to produce the same $\hat\mu$ as the other two binned methods, derived from Eq.~(\ref{eq:muest}), is $k=k_1+k_2-k_1 k_2 / n$.
For the particular example of $k_1=96$ and $k_2=48$, the equivalent number of bars \emph{on} in the single-window likelihood is $k=120$.
Figure~\ref{fig:like1} displays the 1$\sigma$ and 2$\sigma$ confidence intervals defined by the conditions $f(\mu)=1$ and $f(\mu)=4$ respectively. 
The $f(\mu)$  of the profile and integrated likelihoods are very similar and have smaller confidence intervals than the exact likelihood.
The resolution is enhanced with the two approximated methods because they consider the detector timing.

% Saturated counter
The single-window likelihood saturates earlier than the profile one. 
While in the first case the variable $k$ of Eq.~(\ref{eq:kprob}) corresponds to the bars that have a signal over the whole event duration, the $k_i$'s of the profile likelihood refer to a single time bin.
Since this second method spreads the signal over many time bins, $k$ is greater than $k_i$.
Therefore the saturation condition, i.e. all bars \emph{on}, is reached in the single-window likelihood with fewer muons than in the profile method.
Because the integrated and the profile likelihoods rely on the same signal binning, both techniques saturate identically.

The likelihoods corresponding to an event with two time bins, of which the first one is saturated, are displayed in the bottom panel of Fig.~\ref{fig:like1}.  
In this example the profile likelihood imposes a more stringent limit than the single-window method to the number of muons.  
Although the integrated likelihood $f(\mu)$ has a minimum, in practice it only works as a lower bound by imposing a large penalty to small $\mu$'s. 

% The presence of the minimum has to do with the fact that the full likelihood function present a maximum and the integrated likelihood is calculated averaging over all possible likelihood functions.  

\section{Simulations}
\label{sec:simulations}

\begin{figure}[tbh]
\centering
\setlength{\abovecaptionskip}{0pt}
\includegraphics[width=.45\textwidth]{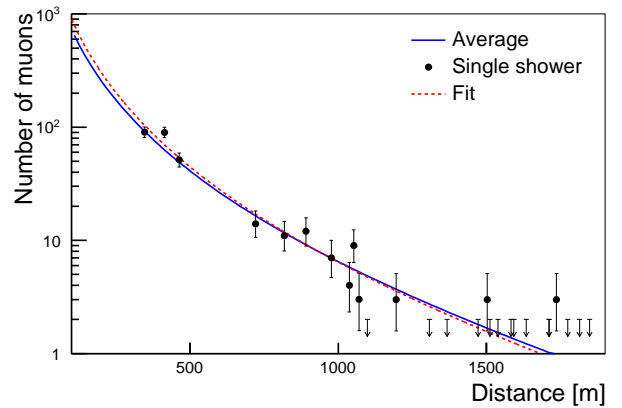}
\caption{Average muon lateral distribution function fitted to the simulated detector data (continuous blue line). 
The fitted data correspond to the average number of muons in the AMIGA detectors calculated with simulations of iron primaries with energy $E=1\,\mathrm{EeV}$ and zenith angle $\theta=30^{\circ}$.
An example of the AMIGA response to a single shower of the same type is also shown for comparison,
together with the corresponding fit of a Kascade-Grande--like muon LDF (dotted red line).
\label{fig:sim1}}
\end{figure}

\begin{figure}[tbh]
\centering
\setlength{\abovecaptionskip}{0pt}
\includegraphics[width=.45\textwidth]{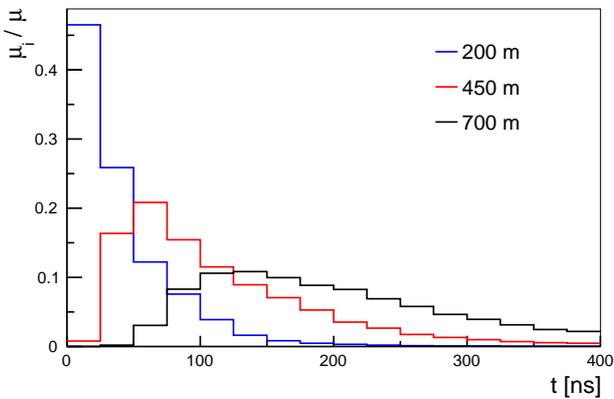}
\caption{Average time histograms at three different distances to the shower axis measured at the shower plane. The bin size of $25\,\mathrm{ns}$ corresponds to the detector time resolution. The histograms show the fraction of muons in each time bin.
\label{fig:sim2}}
\end{figure}

We tested the performance of the different likelihoods with air showers simulated with CORSIKA v7.3700~\cite{Knapp:1998ra} using the high energy hadronic model EPOS-LHC~\cite{Pierog:2013ria}. 
We simulated proton and iron primaries in the energy interval $\log_{10}(E/\textrm{eV}) \in [17.5, 19]$ in steps of $\Delta\log_{10}(E/\textrm{eV}) = 0.25$ for the zenith angles $\theta = 0^\circ$, $30^\circ$, and $45^\circ$.
In the simulations, we applied an algorithm with an optimal statistical thinning of $10^{-6}$ that reduced the number of tracked particles. 
We produced twenty proton and fifteen iron showers for each energy and zenith angle combination.
For each simulation we recorded the number of muons crossing a $30\,\mathrm{m}^2$ area placed $2.5\,\mathrm{m}$ underground as in the AMIGA detectors.
We considered the shielding of the soil by selecting muons with energy greater than $1\,\mathrm{GeV}/\cos\theta$, with $\theta$ the zenith angle of the muon.
We computed the average number of muons as function of the distance to the shower axis, measured at shower plane, over each set of simulated showers and fitted these values with a Kascade-Grande--like muon LDF~\cite{Apel:2010zz}. 
We also produced histograms of the muon arrival times at different core distances.
Figure~\ref{fig:sim1} shows the average LDF and Fig.~\ref{fig:sim2} the arrival time histograms at 3 different distances for $1\,\mathrm{EeV}$ iron showers arriving at $\theta=30^{\circ}$.  
The arrival time histograms show the fraction of particles arriving in $25\,\mathrm{ns}$ time bins with respect to the total number of muons.
We only considered muons above $1\,\mathrm{GeV}/\cos\theta$, the threshold energy required to break through the soil shielding.
The histograms show that muons arrive more spread in time farther away from the shower core. 

For a given energy and zenith angle we sampled each average shower many times varying the azimuth angle and the impact position on the ground.
We adjusted the simulated showers with the single-window, profile and integrated likelihoods.
The integrated likelihood evaluation, involving multidimensional integrals, requires a much larger computational time than the profile likelihood.
Therefore we used different numbers of events with each method; we sampled each shower $1\,000$ times for the integrated likelihood and $10\,000$ times for the other two methods.
Since the processing budget also increases with primary energy, for the integrated likelihood we only reconstructed showers up to $\log_{10}(E/\textrm{eV})=18.5$.

% Data generation
For each sampled event we calculated the distance of the counters to the shower axis.
Then we evaluated the average LDF at each distance to find the number of muons expected in each counter ($\mu$).
Using $\mu$ as a parameter, we sampled the actual number of muons from a Poisson distribution. 
We considered a detector as \emph{untriggered} if it received two or fewer muons.
We obtained the arrival time of each muon by sampling the time distribution histograms and calculated  the number of muons in each $25 \, \mathrm{ns}$ time bin accordingly.
In a second step we randomly distributed the muons across the detector and calculated how many segments were \emph{on}.
The number of strips \emph{on} per time window is the input data to build the likelihood of each detector.
We computed the maximum of this likelihood to obtain an estimator of the muons, $\hat\mu$, in each detector using Eq.~(\ref{eq:muest2}). 
Figure~\ref{fig:sim1} shows, for a single shower, the $\hat\mu$ of each triggered detector.
The untriggered counters are represented in this plot with a down arrow.
For each simulated event we adjusted $\mu$ as function of the core distance with a second Kascade-Grande--like muon LDF.
The energy reconstruction of the events is based on the evaluation of the fitted LDF at an optimal distance ($r_0$) at which the spread of the LDF is minimal~\cite{Newton:2006wy}.
For reasons that will be explained later, it is convenient to make of the LDF value at $r_0$ a parameter of this function ($\mu_0$).
To isolate this parameter we factorised the LDF $\big(\mu(r)\big)$ into a normalisation factor $\mu_0$ and a second function $g(r)$, 
\begin{equation}
\label{eq:LDF}
\mu(r)=\mu_0 \frac{g(r)}{g(r_0)}. 
\end{equation}
\noindent The function $g(r)$, containing the distance dependence, is,
\begin{equation}
\label{eq:fform}
g(r)=\left( \frac{r}{r_1} \right)^{-\alpha} \! \left( 1+\frac{r}{r_1} \right)^{-\beta}%
\! \left( 1+\left( \frac{r}{10 \, r_1}\right)^2 \right)^{-\gamma} \! \! \! \!  , 
\end{equation}

\noindent where $r$ is the distance to the shower axis in the shower front, $\alpha = 0.75$, $r_1 = 320 \,\mathrm{m}$, and $\gamma=2.95$. 
We adjusted $\mu_0$ and the slope $\beta$ by minimising the function,
\begin{equation}
\label{eq:Lfit}
-2 \, \ln L_{fit}(\mu_0,\beta) = -2 \, \sum_i \ln\lambda_i(\mu(r_i,\mu_0,\beta)),
\end{equation}

\noindent where the sum runs over the detectors. For the $i$-th counter, $\lambda_i$ is the function introduced in section~\ref{subsec:proflike}, and $r_i$ is the core distance.
The input data of the fit are, through the $\lambda_i$ functions, the number of strips \emph{on} per time window in each counter.
For untriggered counters we used a Poisson likelihood, setting an upper limit to the number of muons allowed in the LDF fit as in Ref.~\cite{Supanitsky:2008dx}. 
Figure~\ref{fig:sim1} shows the fit of the detector data simulated for a single shower using the profile likelihood.

\section{Reconstruction performance}
\label{sec:performance}

In this section we evaluate the performance of the reconstructions using the single-window, profile, and integrated likelihoods.
For this assessment we compared the bias and the fluctuations of the $\mu_0$ inferred with each method.
In addition to the properties of this point estimator, we also look at the size of the $\mu_0$ confidence intervals derived from the LDF reconstructions.
For brevity we only show the results of iron primaries at $\theta = 30^\circ$; the proton showers and the other simulated zenith angles have similar outcomes.

\subsection{Saturation}

\begin{figure} [tbh]
\centering
\setlength{\abovecaptionskip}{0pt}
\includegraphics[width=.45\textwidth]{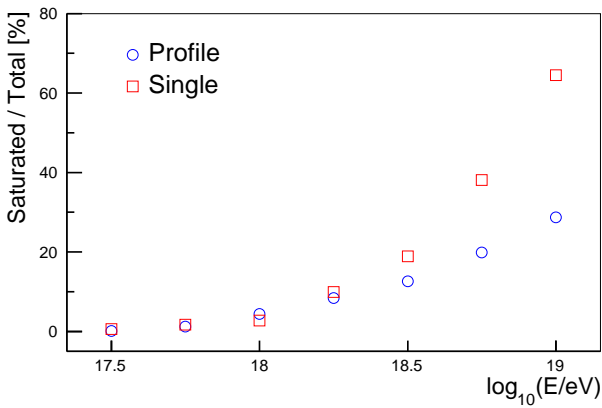}
\caption{Fraction of saturated events for iron primaries at $\theta = 30^\circ$. The integrated likelihood has the same saturation as the profile method. The detector saturates more with the single-window likelihood than with the other two methods.
\label{fig:satAmiga}} 
\end{figure}

The fraction of saturated events increases with the primary energy as the signal deposited in the detectors raises.
Given that signals are spread in many time bins, detectors saturate less with the profile and the integrated likelihoods than with the single-window method. 
Figure~\ref{fig:satAmiga} displays the fraction of saturated events with respect to the total number of simulated events as function of energy for the profile and single-window reconstructions.
The integrated likelihood, using the same time window size, has the same saturation as the profile method.
Since $40\%$ of the events saturate at $\log_{10}(E/\textrm{eV})=18.75$, we cut the analysis of the single-window likelihood at this energy.  

% Comments on saturation and electromagnetic contamination
For the comparisons we only selected events which have all detectors free of saturation.
We excluded saturated events because their shower size parameters are reconstructed with a significant bias~\cite{Ravignani:2014jza}.
Given the steepness of the lateral distribution of shower particles, saturation happens mainly in detectors close to the core.
In these detectors the muon signal may also be contaminated by electromagnetic particles and hadrons. 
Preliminary simulations of AMIGA show that this contamination is below 1\% at $100\,\mathrm{m}$ from the shower core~(J. M. Figueira, personal communication, 21 April 2016).
This distance is less than the average distance of the nearest detector to the shower core, which is $230\,\mathrm{m}$ according to Ref.~\cite{Supanitsky:2008dx}.
More detailed simulations are currently under way to study the punch trough of electromagnetic particles.
These simulations will confirm whether the punch trough can be neglected or not.
If the contamination effect has to be considered, the current likelihood model will have to be updated accordingly.

\subsection{Optimal distance}

\begin{figure} [tbh]
\centering
\setlength{\abovecaptionskip}{0pt}
\includegraphics[width=.45\textwidth]{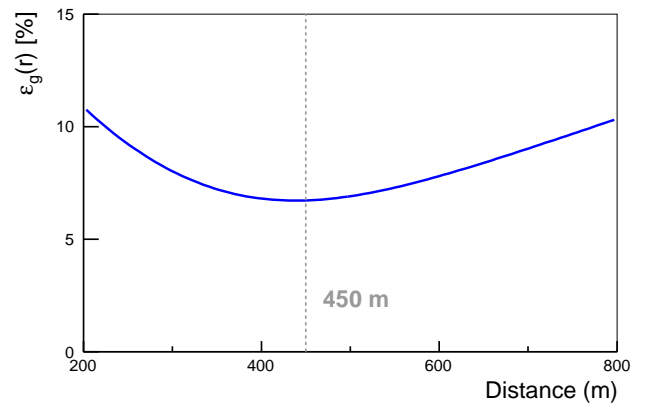}
\caption{Relative standard deviation of the lateral distribution function reconstructed using the profile likelihood. The curve corresponds to a global average calculated using all simulated showers. A minimum is reached close to $450\,\mathrm{m}$.
\label{fig:rmin}}
\end{figure}

The statistical fluctuations of the detector data are caused by the combined contributions of the finite number of muons and the detector segmentation. 
These variations propagate during the fit to the estimated LDF parameters, introducing fluctuations in the reconstructed LDF. 
We evaluated the standard deviation of the fitted LDF as function of the core distance ($\sigma(r)$) using
\begin{equation}
\label{sigma}
\sigma(r)^2 = \frac{\sum_{i=1}^N  (\mu_i(r) - \bar{\mu}(r))^2}{N-1},  
\end{equation}
\noindent where $N$ is the number of simulations, $\mu_i$ corresponds to the $i$-th reconstructed LDF, and $\bar{\mu}$ to the $\mu_i$'s average. 
We calculated the relative standard deviation of the LDF ($\varepsilon(r)$) dividing $\sigma(r)$ by $\bar{\mu}$.
The function $\varepsilon(r)$ represents the accuracy with which the array reconstructs the muon number at different distances.
We derived first an $\varepsilon(r)$ for each simulated primary type, shower energy, and zenith angle, respectively. 
Afterwards we added these functions in quadrature to obtain a global resolution $\varepsilon_{g}(r)$. 
Figure~\ref{fig:rmin} shows the $\varepsilon_g(r)$ corresponding to reconstructions with the profile likelihood. 
The function $\varepsilon_g(r)$ reaches a minimum close to $r_0=450\,\mathrm{m}$. 
This is, therefore, the optimal distance to measure the number of muons with AMIGA.  
The value of the reconstructed LDF at $r_0$ is taken as the shower size estimator (${\hat\mu}(450)$).

% Justification of a single optimal distance
The optimal distance of a segmented detector array like AMIGA depends on the primary type, energy, and zenith angle.
However the $\varepsilon(r)$ value at the optimal distance of each specific shower type and the corresponding value at $r_0=450\,\mathrm{m}$ differed in less than $0.5\%$ in all simulations.  
Therefore the convenience of adopting a single optimal distance for all events outweighs any resolution loss introduced by not using a different optimal distance for each shower type.    
In addition, the optimal distances of the single-window and integrated likelihoods are also close to $r_0=450\,\mathrm{m}$.
So, to ease the comparison between the different methods, we adopted the same $r_0$ for all of them.

\begin{figure} [tbh]
\centering
\setlength{\abovecaptionskip}{0pt}
\includegraphics[width=.45\textwidth]{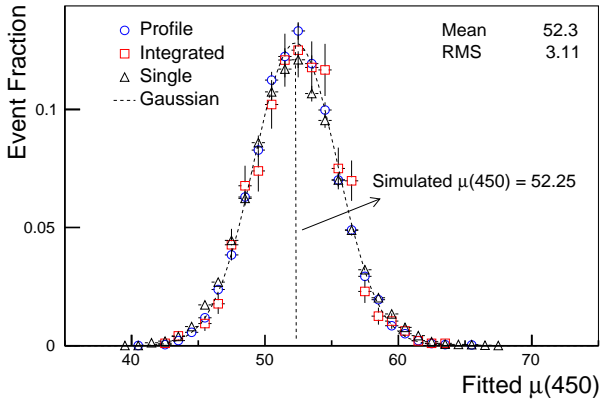}
\caption{Distribution of the reconstructed number of muons at 450 m from the shower axis using the profile, integrated, and single-window likelihoods. The data of this and the following plots correspond to simulations of $1\,\mathrm{EeV}$ iron showers at $\theta = 30^\circ$. The histogram mean matches the simulated $\mu(450)$.
\label{fig:mu450}}
\end{figure}

Given the fluctuations in the detector signals, the fitted ${\hat\mu}(450)$ varies across reconstructions of the same shower.
Figure~\ref{fig:mu450} shows histograms of the ${\hat\mu}(450)$ reconstructed with the profile, integrated, and single-window likelihoods for $1\,\mathrm{EeV}$ iron showers arriving at $\theta = 30^\circ$.
The three histograms coincide within statistical uncertainties. 
Since ten times less reconstructions were run for the integrated likelihood, its data have larger error bars than the other two methods.
The plot also displays a Gaussian distribution parametrised with the mean and the standard deviation of the profile likelihood histogram.
The distributions of ${\hat\mu}(450)$ are well described by the Gaussian.
In this example the ${\hat\mu}(450)$ distributions are unbiased, i.e. the histogram means match the $\mu(450)$ of the input LDF. 
For the three considered likelihoods the relative standard deviation of ${\hat\mu}(450)$ is close $\varepsilon(450)=6\%$.
In the shown example, the ${\hat\mu}(450)$ distributions of the three likelihoods are similar because the shower $\mu(450)$ is much smaller than the $192$ segments of the AMIGA detector.

\subsection{Bias}

\begin{figure} [tbh]
\centering
\setlength{\abovecaptionskip}{0pt}
\includegraphics[width=.45\textwidth]{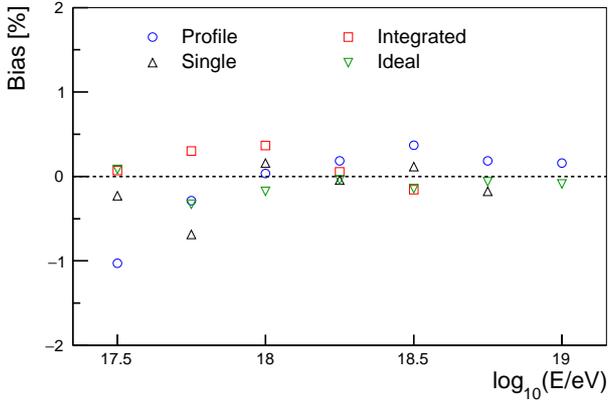}
\caption{Relative bias of number of muons at $450\,\mathrm{m}$ from the shower core. Reconstructions with the profile, integrated, and single-window likelihoods together with the case of an ideal particle counter are shown. All observed biases are of the order of 1\% or less. \label{fig:bias}}
\end{figure}

%Bias
The comparison of the input $\mu(450)$ and the corresponding value fitted afterwards to the simulated data is a valuable method to assess the reconstruction performance. 
We estimated the bias as the difference between the average $\bar{\mu}(450)$ calculated over the reconstructions and the input $\mu(450)$.
As the reconstructed ${\hat\mu}(450)$ changes according to the likelihood applied in the LDF fit, the ${\hat\mu}(450)$ bias can also vary among the different methods. 
Figure \ref{fig:bias} shows their relative biases, calculated as the bias over $\mu(450)$, versus energy.
The case of an ideal detector, that counts particles without any pile-up effect, is also included in the comparison. 
The likelihood used for this detector is the Poissonian,
\begin{equation}
\label{eq:lpoisson}
L(\mu) = e^{-\mu} \, \frac{\mu^k}{k!}.
\end{equation}
\noindent where $k$ is the number of counted particles. All observed biases are of the order of 1\% or less, the four methods can be considered as unbiased.       

\subsection{Standard deviation}
%context
The second quantity used to evaluate the reconstruction performance is the standard deviation of the ${\hat\mu}(450)$ reconstructed in the LDF fit ($\sigma(450)$).
The $\sigma(450)$ measures the fluctuations of the ${\hat\mu}(450)$ fitted for a single event around the mean calculated over all events.
Since the combination of a small bias and a low standard deviation allows for a good estimation of $\mu(450)$ using the data from a single event, a small $\sigma(450)$ is a desirable property of the reconstructed ${\hat\mu}(450)$.  

\begin{figure} [tbh]
\centering
\setlength{\abovecaptionskip}{0pt}
\includegraphics[width=.45\textwidth]{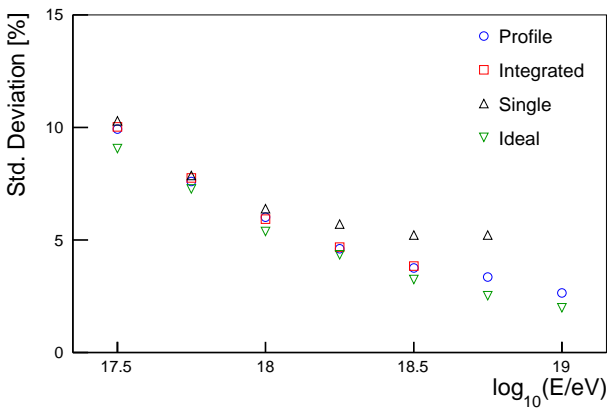}
\caption{Relative standard deviation of the muon density at $450\,\mathrm{m}$ from the shower core. The uncertainties of the four shown methods are similar up to $\log_{10}(E/\textrm{eV})=18$; at higher energies the reconstruction with the single-window likelihood has less resolution than the other three cases. \label{fig:sigma}}
\end{figure}

%results
For the four evaluated likelihoods, we estimated the $\sigma(450)$ relative to $\mu(450)$ (i.e $\varepsilon(450)$).
Figure~\ref{fig:sigma} shows the corresponding $\varepsilon(450)$ as function of energy for iron showers at $\theta = 30^\circ$.
The $\varepsilon(450)$ improves with energy because showers contain more muons; with more particles more detectors are triggered and counters have higher signals.
The $\varepsilon(450)$ calculated with the four methods is similar up to $1\,\mathrm{EeV}$. 
At higher energies the profile reconstruction has a better resolution than the single-window one. 
With the single-window likelihood the resolution flattens as muons start to pile up in the counters. 
The effect is more noticeable at high energy, when there are more muons and therefore they accumulate more. 
On the other hand, by using the profile and integrated likelihoods muons distribute over many time windows, so there are fewer muons per time bin than in the single-window case.
The $\varepsilon(450)$ of the integrated and profile likelihoods are close up to $\log_{10}(E/\textrm{eV}) = 18.5$, the highest simulated energy for the integrated likelihood.
The ideal counter sets a lower bound to the $\varepsilon(450)$ achievable with an AMIGA like array of $30\,\mathrm{m^2}$ detectors. 
In the considered energy range, the $\varepsilon(450)$ of the profile likelihood is almost similar to this best case scenario.

\subsection{Coverage}

% In defense of coverage
The bias and standard deviation are properties of point estimators like, in this case, ${\hat\mu}(450)$. 
On the other hand, coverage is the main measure of the confidence interval quality.
For each event the 1$\sigma$ errors of the LDF normalisation ${\hat\mu}(450)$ and the slope parameter $\beta$ are calculated during the reconstruction by setting  $-2 \, \ln L_{fit}$ in Eq.~(\ref{eq:Lfit}) equal to one.   
We parametrised the LDF with ${\hat\mu}(450)$ in Eq.~(\ref{eq:LDF}) to obtain its confidence interval directly from the fit procedure.
The coverage of a confidence interval is defined as the probability it contains the true value of the estimated parameter.
For example, the coverage of the 1$\sigma$ interval of a Gaussian distribution is $0.68$. 
In the more general case of a distribution approximately Gaussian the coverage is expected to be close to this value. 
If the data errors are underestimated, or conversely the likelihood is too narrow, the coverage of the confidence intervals derived from the fit can be significantly lower than the Gaussian value.
This property is equivalent to the high $\chi^2$ produced in a fit when data errors are underestimated.
In this sense, coverage is another way of measuring the goodness of a fit. 
But while the $\chi^2$ usually refers to a single fit, coverage quantifies quality over many events.

\begin{figure} [tbh]
\centering
\setlength{\abovecaptionskip}{0pt}
\includegraphics[width=.45\textwidth]{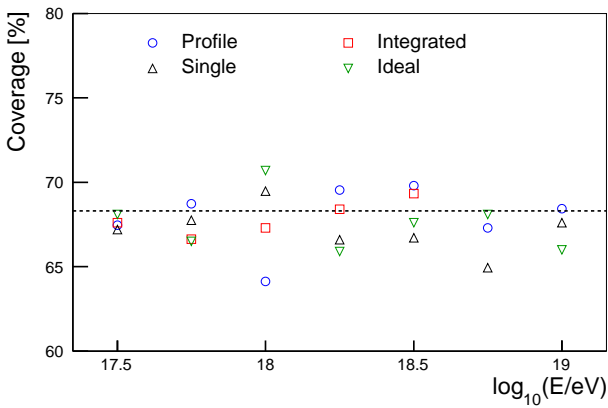}
\caption{Coverage of the 1$\sigma$ confidence interval of $\mu(450)$. The dotted line shows the coverage of a Gaussian distribution. The coverage of the four reconstruction methods are approximately similar to each other and to the Gaussian value. \label{fig:coverAmiga}}
\end{figure}

% Measured coverage
We estimated the coverage of the $\mu(450)$ confidence intervals as the fraction of reconstructed events that included, within the mentioned intervals, the input value used in the LDF simulations. 
Figure~\ref{fig:coverAmiga} shows the coverage of the reconstructions of an iron primary at
$\theta = 30^\circ$ at different energies.
This plot also shows the coverage of the 1$\sigma$ interval corresponding to a Gaussian distribution. 
The coverage of all reconstructions are close to each other and to the Gaussian reference. 

\section{Conclusions}
\label{sec:conclusions}

% Two improvements
We introduced two different methods to reconstruct the lateral distribution function of air shower muons: the profile and the integrated likelihoods.
Both likelihoods extend a previous approach by considering the detector timing.
Although we applied the likelihoods to a specific cosmic ray detector, they can be used for any kind of segmented particle counters with time resolution.
We found an optimal distance of $450\,\mathrm{m}$ to measure the shower size parameter in a triangular array with $750\,\mathrm{m}$ between detectors. 
The new likelihoods improve the reconstruction in two aspects.
Firstly, by raising the number of muons a detector can handle before saturating, more events can be reconstructed. 
The recovery is more significant close to $10\,\mathrm{EeV}$, the upper limit of the considered energy range, a region where events are usually scarce.
Secondly, we reduced the statistical fluctuations of the parameter that measures the shower size from $1\,\mathrm{EeV}$ upwards.  
This decrease allows for a more powerful discrimination between different primary masses based on the number of muons.
By comparing to an ideal muon counter, we established that the resolutions achieved with the new likelihoods are close to the lower bound given the detector size and spacing. 
We also showed that the approximations introduced for the profile and integrated likelihoods do not bias the reconstructed shower size parameter and kept the coverage of its 1$\sigma$ confidence interval close to the expected Gaussian nominal value. 

% Why profile
The shower size parameters reconstructed with the integrated and the profile likelihoods are very similar.
Nevertheless the profile likelihood is the preferred reconstruction method given the much shorter time it takes to process the data.
The correspondence between the profile and the integrated likelihood results, shows the robustness of these techniques to reconstruct the muon lateral distribution with an array of segmented counters.

\appendix

\section{Integrated likelihood multiplicity}
\label{App}

In order to prove Eq.~(\ref{IntLm}) let us write Eq.~(\ref{eq:IntL}) in the following way,
\begin{equation}
\label{IntLa1}
L_I (\mu) = \int_0^1 dp_1 \cdots \int_0^1 dp_N \prod_{i=1}^N f(p_i,k_i)\  
\delta\left(\sum_{i=1}^N p_i - 1 \right),   
\end{equation} 
\noindent where $f(p_i,k_i)=\exp(-\mu\ p_i) \left( \exp(\mu\ p_i/n)-1 \right)^{k_i}$. Then, if there are $m$ time intervals that have the same $k$, it is possible to choose the first $m$ values of $i$ such that $k_1=\cdots=k_m=k$. Let us consider the integral,
\begin{equation}
\label{intx}
\int_0^1 dx\ \delta\left(x-\sum_{i=1}^m p_i \right) = \Theta\left(1-\sum_{i=1}^m p_i\right) = 1,
\end{equation}
\noindent where $\Theta(x)=1$ if $x \geq 0$ and $\Theta(x)=0$ if $x<0$. Here it is used that $\sum_{i=1}^m p_i \leq 1$. If the change of variable $x=m\ \xi$ is considered, Eq.~(\ref{intx}) is written as,
\begin{equation}
\label{intxi}
m \int_0^{1/m} d\xi\ \delta\left(m\ \xi-\sum_{i=1}^m p_i \right) = 1.
\end{equation}

\noindent Therefore, inserting Eq.~(\ref{intxi}) in Eq.~(\ref{IntLa1}) and integrating over $p_{m}$ the following expression is obtained,
\begin{equation}
\label{IntLa2}
\begin{aligned}
L_I (\mu) = & \int_0^1 dp_{m+1} \cdots \int_0^1 dp_N \int_0^{1/m} d\xi\ \prod_{i=m+1}^N f(p_i,k_i) \\  
& \delta\left(m\ \xi + \sum_{i=m+1}^N p_i - 1 \right)\ g(\xi,k,m),   
\end{aligned}
\end{equation} 
\noindent where 
\begin{equation}
\label{gdef}
\begin{aligned}
g(\xi,k,m) =& m \int_0^1 dp_{1} \cdots \int_0^1 dp_{m-1} \prod_{i=1}^{m-1} f(p_i,k)\\ 
&f\left(m\ \xi-\sum_{i=1}^{m-1} p_i,k\right)\  
\Theta\left(m\ \xi - \sum_{i=1}^{m-1} p_i \right),   
\end{aligned}
\end{equation} 

\noindent The integral in Eq.~(\ref{gdef}) cannot be analytically solved, then an approximated expression is obtained. For that purpose, let us 
consider the function,
\begin{eqnarray}
h(p_1,...,p_{m-1}) \!\!\! &=& \!\!\! \ln\left[  \prod_{i=1}^{m-1} f(p_i,k)\ f\left(m\ \xi-\sum_{i=1}^{m-1} p_i,k\right) \right], \nonumber \\
\!\!\! &=& \!\!\! \sum_{i=1}^{m-1} \ln f(p_i,k)\ +\ln f\left(m\ \xi-\sum_{i=1}^{m-1} p_i,k\right). \nonumber
\end{eqnarray} 
\noindent It is easy to see that,
\begin{equation}
\frac{\partial h}{\partial p_j} (p_1=\xi,...,p_{m-1}=\xi) = 0,
\end{equation}
\noindent which means that the vector $\vec{p}=(\xi, ... , \xi)$ is an extreme of $h$. Note that this property does not depend on the specific form of $f$. The elements of the Hessian matrix evaluated in this vector are given by,
\begin{equation}
\begin{aligned}
\frac{\partial^2 h}{\partial p_i \partial p_j} (p_1=\xi,...,p_{m-1}=\xi) =& -\frac{\exp(\mu\ \xi/n)}{(\exp(\mu\ \xi/n)-1)^2} \\
& \times \frac{k \mu^2}{n^2} (1+\delta_{ij}),
\end{aligned}
\end{equation}
\noindent where $\delta_{ij}$ is the Kronecker delta ($\delta_{ii}=1$ and $\delta_{ij}=0$ for $i \neq j$). Note that the diagonal elements of the Hessian matrix are negative, which means that $\vec{p}=(\xi, ... , \xi)$ is a maximum. Considering just the zero order of the Taylor expansion of $h$ at $\vec{p}=(\xi, ... , \xi)$ the following expression for $g$ is obtained,
\begin{eqnarray}
g(\xi,k,m) \!\!\! &\cong & \!\!\! m\ f(\xi,k)^m \int_0^1 dp_{1} \cdots \int_0^1 dp_{m-1} \nonumber \\
\!\!\! && \!\!\!\Theta\left(m\ \xi - \sum_{i=1}^{m-1} p_i \right) \nonumber \\
\label{gapp}
\!\!\! &\cong & \!\!\! \frac{m^m}{(m-1)!}\ \xi^{m-1} f(\xi,k)^m.
\end{eqnarray}

\noindent Then, inserting Eq.~(\ref{gapp}) in Eq.~(\ref{IntLa2}) we obtain,
\begin{equation}
\label{IntLa3}
\begin{aligned}
L_I (\mu) \cong& \frac{m^m}{(m-1)!} \int_0^1 dp_{m+1} \cdots \int_0^1 dp_N \int_0^{1/m} d\xi\ \prod_{i=m+1}^N f(p_i,k_i)\\  
& f(\xi,k)^m\ \xi^{m-1} \ \delta\left(m\ \xi + \sum_{i=m+1}^N p_i - 1 \right).
\end{aligned}
\end{equation}    
\noindent Therefore, Eq.~(\ref{IntLm}) is straightforwardly obtained from Eq.~(\ref{IntLa3}). 

\section*{Acknowledgements}
The authors have greatly benefited from discussions with several colleagues from the Pierre Auger Collaboration, of which they are members.
We especially thank R. Clay for a careful review of the manuscript.  
A. D. Supanitsky and D. Melo are members of the Carrera del Investigador Cient\'{\i}fico of CONICET, Argentina. 
This work was partially funded by PIP 114-201101-00360 (CONICET) and PICT 2013-1934 (ANPCyT).

\end{document}